\documentclass[reprint,showpacs,amssymb,amsmath,aps,pra]{revtex4-1}
\usepackage{graphics}
\usepackage{bbold}
\usepackage{graphicx}
\usepackage{bm}
\usepackage{amsmath}
\usepackage{amssymb}
\begin{document}

\title{Quantum Discord and entropic measures of quantum correlations: Optimization 
and behavior in finite $XY$ spin chains}
\author{N.\ Canosa$^1$, M.\ Cerezo$^1$, N.\ Gigena$^1$, R.\ Rossignoli$^{1,2}$}
\affiliation{$^1$Departamento de F\'{\i}sica-IFLP,
Universidad Nacional de La Plata \\ C.C. 67, La Plata (1900), Argentina\\
$^2$Comisi\'on de Investigaciones Cient\'{\i}ficas (CIC), La Plata (1900), Argentina}

\begin{abstract}
We discuss a generalization of the conditional entropy 
and one-way information deficit in quantum systems, based on general 
entropic forms. The formalism allows to consider simple entropic forms 
for which a closed evaluation of the associated optimization 
problem in qudit-qubit systems is shown to become feasible, allowing to approximate 
that of the quantum discord. As application, we examine quantum correlations of 
spin pairs in the exact ground state of finite $XY$ spin chains in a magnetic field 
through the quantum discord and information deficit. While these quantities show a 
similar behavior, their optimizing measurements exhibit significant differences, 
which can be understood and predicted through the previous approximations. 
The remarkable behavior of these quantities in the vicinity of transverse and 
non-transverse factorizing fields is also discussed.
\end{abstract}
\maketitle

\section{Introduction
\label{sec1}}

Non-classical correlations in mixed states of composite 
quantum systems have attracted strong attention in recent years \cite{Mo.12,ABC.16}. In pure
states they can be identified with entanglement 
\cite{Sch.95-1,Sch.95-2,NC.00,HR.07} and are essential for quantum teleportation \cite{CHBe.93} 
 and for achieving exponential speed-up in pure state based quantum algorithms 
\cite{JL.03,GV.03}. However, in the case of mixed states  
it is now well known that separable states, defined in general as convex mixtures of product states 
 \cite{RW.89}, i.e. those which can then be created by local operations and classical communication 
\cite{NC.00,RW.89}, may still exhibit non-classical features, such as a non-zero value of the quantum 
 discord \cite{OZ.01,HV.01,HV.03,Zu.03}. The latter is defined as the difference between two distinct  quantum extensions of the classical mutual information or conditional entropy,  becoming zero for  classically correlated states and reducing to the entanglement entropy for pure states.  A finite quantum discord is also present in the mixed state based quantum algorithm of Knill and Laflamme \cite{KL.98}, as shown in \cite{DSC.08},  which achieves an exponential speed-up over classical algorithms without substantial entanglement \cite{DFC.05}. 
This fact triggered the interest not only  in the quantum discord and its fundamental properties 
\cite{KW.04,Dat.09,SL.09,Mo.10,FA.10,FCOC.11} but also in other related measures with similar features \cite{Mo.12,ABC.16}, which include among others 
the one-way information deficit \cite{HHH.05,HO.02,SKB.11,Mo.12}, 
the geometric discord \cite{DVB.10},  the generalized entropic measures introduced in 
\cite{RCC.10,RCC.11} (which contain the previous ones as particular cases),  the
local quantum uncertainty \cite{GTA.13,L.12}, the trace distance discord
\cite{POS.13, HF.13,NPA.13} and more recently coherence based measures \cite{BCP.14,MS.14,ABC.16}.  Besides, various operational interpretations 
of the quantum discord and  other related  measures  have been  provided
\cite{Mo.12, Dat.09, SKB.11, MD.11,CC.11,PG.11,GA.12,TG.12}. It is worth
mentioning, however, that most of  these measures require the determination of  
an optimizing local measurement, which makes their evaluation
difficult in a general situation  (shown to be NP-complete \cite{IW.14}). 

Interacting spin chains provide
a useful scenario for studying  the previous measures and their behavior in the vicinity of critical points \cite{Dd.08,SA.09,MG.10,WR.10,CRC.10,WR.10b,WR.11,BQL.11,YH.11,CCR.12,CCR.13,Mo.12}. 
In general, ground states of interacting spin chains are strongly entangled states,
implying that the state of a reduced spin pair or group of spins will typically be a mixed state.  Hence, for these subsystems differences between discord type
measures  and entanglement will  arise already at zero temperature.  

In this chapter  we first briefly review in section \ref{sec2} the quantum discord and the associated local measurement dependent conditional entropy on which it is based. We then discuss the consistent generalization of this entropy 
to general entropic forms \cite{GR.14,GR.14b}. This extension enables in particular the consideration of simple forms which allow an analytic solution of the associated optimization problem, 
i.e., that of determining the local measurement leading to the lowest conditional entropy, 
for general mixed states of qudit-qubit systems \cite{GR.14,GR.14b}. The solution is given  in terms of an eigenvalue equation which admits a simple geometrical picture \cite{GR.14}. We then examine the generalized 
information deficit \cite{RCC.10}, based on general entropic forms, which contains the standard one-way information deficit \cite{HHH.05,HO.02,SKB.11} as a particular case, together with its exact minimization for simple quadratic entropic forms for general states of  qudit-qubit systems 
\cite{DVB.10,RCC.11}. 

In section \ref{sec3} we will analyze the exact behavior of the quantum discord and the information deficit associated with spin pairs in the ground state of finite $XY$ spin $1/2$ chains immersed in a magnetic field. We will show that while their behavior is quite similar, significant differences do arise in their corresponding minimizing measurements, which can be correctly predicted and understood by the approximations based on simple entropic forms. A remarkable effect in these chains is the possibility of exhibiting a completely separable exact ground state at a factorizing field. The existence of a factorizing field was first discussed in Ref.\ \cite{K.82} 
and its properties together with the general conditions for its existence at transverse  fields
were  analyzed in \cite{R.04, R.05,A.06, B.07, RCM.08, GAI.08, GAI.09, RCM.09,CRM.10,CRGB.13}. The 
transverse factorizing field actually corresponds to the last ground state parity transition
\cite{RCM.08, RCM.09,CRM.10}, and accordingly, it will be shown that in finite chains the quantum discord and information deficit exhibit full range in its vicinity, with an appreciable finite limit value at this field. We will also discuss the behavior for a non-transverse field \cite{CRC.15}, which will differ from the previous one due to the broken spin parity symmetry. Conclusions are finally given in section \ref{sec4}. 

\section{Formalism 
\label{sec2}}
We first describe the main features of the quantum discord and the generalized conditional 
entropy and information deficit, together with some analytic results for general states of 
qudit-qubit systems. 
\subsection{Quantum Discord and conditional entropy}
\label{subsec:1}

Let us start with the well known quantum discord, introduced in
\cite{OZ.01,HV.01}. For a bipartite quantum  system $A+B$ initially in a state
$\rho_{AB}$, it can be defined as the minimum difference between two distinct
quantum versions of the mutual information, or equivalently,  of the  conditional
entropy:
\begin{eqnarray}D(A|B)&=&\mathop{\rm Min}_{M_B}[I(A,B)-I(A,B_{M_B})]\label{D1}\\
 &=&\mathop{\rm Min}_{M_B} S(A|B_{M_B})-S(A|B)\,\label{D2}\end{eqnarray}
where the minimization is over all possible local POVM measurements \cite{NC.00} $M_B$ on
$B$, characterized by a set of operators $M_j=I_A\otimes M_{jB}$ satisfying
$\sum_j M_j^\dagger M_j=I_A\otimes I_B$. Here $I(A,B)=S(\rho_A)-S(A|B)$ represents the
quantum mutual information \cite{Ww.78} before the measurement and
$I(A,B_{M_B})=S(\rho_A)-S(A|B_{M_B})$ a measurement dependent mutual
information, with
\begin{eqnarray}S(A|B)&=&S(\rho_{AB})-S(\rho_B) \label{SAcB1}\\
 S(A|B_{M_B})&=&\sum_ j p_j S(\rho_{A/j})\label{SAcB2}\end{eqnarray}
the corresponding conditional entropies, where
\begin{equation}\rho_{A/j}=p_j^{-1}{\rm Tr}_B\,\rho_{AB}M_j^\dagger M_j
\end{equation}
is the reduced state of $A$ after outcome $j$ at $B$, with $p_j={\rm
Tr}\,\rho_{AB}M_j^\dagger M_j$ the probability of such outcome, and
$S(\rho)=-{\rm Tr}\,\rho\log_2\rho$  the von Neumann entropy. In the case of
complete local {\it projective} measurements $M_j=P_j=I_A\otimes P_{jB}$, with
$P_{jB}\equiv |j_B\rangle\langle j_B|$ one-dimensional orthogonal projectors
($P_{jB}P_{kB}=\delta_{jk}P_{jB}$),   then
\begin{equation}S(A|B_{M_B})=S(\rho'_{AB})-S(\rho'_{B})\label{Pj}\end{equation}
where $\rho'_{AB}$ is the joint state after the (unread) local measurement,
\begin{equation}\rho'_{AB}=\sum_j P_j\rho_{AB}P_j=\sum_j p_j \rho_{A/j}\otimes P_{jB}
\label{rhop}
\end{equation}
and $\rho'_B={\rm Tr}_{A}\rho'_{AB}=\sum_j p_jP_{jB}$ the ensuing state of $B$.

As is well known, the mutual information $I(A,B)$ is a measure of  all
correlations between subsystems $A$ and $B$, being  non-negative and vanishing
just for product states $\rho_{AB}=\rho_A\otimes\rho_B$ \cite{Ww.78}. Eqs.\
(\ref{D1})--(\ref{D2}) can then be regarded as the difference between all
correlations present in the original state and the classical correlations that
remain  after the local measurement on $B$, measuring then the quantum
correlations. Accordingly, $D(A|B)$ is always non-negative \cite{OZ.01,HV.01}, a property which
stems from the concavity of the {\it conditional} von Neumann entropy $S(A|B)$
\cite{Ww.78}. It vanishes just for semi-quantum states $\rho_{AB}$, which are
already of the form (\ref{rhop}) and which then remain invariant under the
local measurement determined by the projectors $P_{jB}$. The quantum discord is
then non-zero not only in entangled states but also in most separable mixed
states, i.e., those  not of the form (\ref{rhop}) (and hence not diagonal in a
conditional product basis $\{|i_A^j\rangle|j_B\rangle\}$). For pure states
($\rho_{AB}^2=\rho_{AB}$) it reduces to the entanglement entropy
$S(\rho_A)=S(\rho_B)$ of the system, as $S(A|B_{M_B})=0$ for {\it any}
measurement based on rank one projectors.

We remark that in the general case, the minimum in Eq.\ (\ref{D2}) is always
reached for measurements based on rank one projectors $M_{jB}\propto P_{jB}$, not
necessarily orthogonal \cite{Mo.12,GR.14,GR.14b}, with a minimization based on
standard projective measurements, Eq.\ (\ref{Pj}), providing normally a good
approximation. Nevertheless, the minimization in Eq. (\ref{D2}) is in general
still difficult, being in fact an NP-complete problem \cite{IW.14}.

\subsection{Generalized conditional entropy after a local measurement}
Due the previous difficulty, and in order to obtain a more clear picture of the
optimization problem associated with the quantum discord, it is convenient to
consider more simple entropic forms, which may enable an easier evaluation of
the minimum conditional entropy. We then consider first the generalized
conditional entropy \cite{GR.14,GR.14b}
\begin{equation} S_f(A|B_{M_B})=\sum_j p_j S_f(\rho_{A/j})\label{Sfc}\end{equation}
where $S_f(\rho)={\rm Tr}\,f(\rho)$ is a generalized trace form entropy \cite{Ww.78,CR.02}. Here
$f:[0,1]\rightarrow \mathbb{R}$ is a smooth strictly concave function
satisfying $f(0)=f(1)=0$, such that $S_f(\rho)\geq 0$,  with $S_f(\rho)=0$ just
for pure states. Concavity of $f$ implies, for $S_f(A)\equiv S_f(\rho_A)$
\cite{GR.14},
\[S_f(A)\geq S_f(A|B_{M_B})\]
so that the average conditional mixedness of $A$ after measurement is never
greater than the original mixedness, irrespective of the measure $S_f$ used to
quantify it. Moreover, the minimum of $S_f(A|M_B)$ is also always  reached for
rank one projectors $M_{jB}\propto P_{jB}$ \cite{GR.14b}, as in the von Neumann case. 
Hence, these properties remain valid for general concave functions $f$.

In particular, we may consider simple entropic forms, like the quadratic
entropy
\begin{equation}S_2(\rho)=2[1-{\rm Tr}\,\rho^2]\label{S2}\end{equation}
which follows from $f_2(\rho)=2\rho(1-\rho)$ and is also known as linear
entropy since it corresponds to the linear approximation $-\rho\ln \rho\approx
\rho(1-\rho)$. It is a particular case of the Tsallis entropies \cite{TS.09}
$S_q(\rho)=\frac{1-{\rm Tr}\,\rho^q}{1-2^{1-q}}$, obtained for
$f_q(\rho)\propto \rho-\rho^q$, $q>0$, which approach the von Neumann entropy
for $q\rightarrow 1$ (we set $S_f(\rho)=1$ for a maximally mixed single qubit
state).

Eq.\ (\ref{S2}) is just a linear function of the purity ${\rm Tr}\,\rho^2$ and
does not require the explicit knowledge of the eigenvalues of $\rho$, thus
enabling an easier evaluation, both theoretically and experimentally
\cite{F.02,F.02b,F.02c}. For instance, writing a general mixed state of a system
with Hilbert space dimension $d$ as
\begin{equation}\rho=\frac{1}{d}(I+\bm{r}\cdot\bm{\sigma})\end{equation}
where $\bm{\sigma}=(\sigma_1,\ldots,\sigma_{d^2-1})$ is an orthogonal basis for
traceless operators in the system (${\rm Tr}\,{\sigma}_\mu=0$, ${\rm
Tr}\,{\sigma}_\mu\sigma_\nu=d\delta_{\mu\nu}$), implying $\bm{r}={\rm
Tr}\,\rho\bm{\sigma}=\langle\bm{\sigma}\rangle$, we  obtain the explicit
expression
\begin{equation}S_2(\rho)=\frac{2}{d}(d-1-|\bm{r}|^2)\,.\label{S2r}\end{equation}
Eq.\ (\ref{S2r}) shows that $|\bm{r}|^2\leq d-1$, with $|\bm{r}|^2=d-1$ just
for pure states.

\subsection{The qudit-qubit case }
Let us now consider a composite system where $A$ is a system with Hilbert space
dimension $d_A$ and  $B$ a single qubit. Denoting with $\bm{\sigma}_A$ an
orthogonal basis for operators in $A$ and $\bm{\sigma}_B\equiv\bm{\sigma}$ the Pauli matrices of
$B$, a general state of this system can be  written as \cite{GR.14}
\begin{equation}\rho_{AB}=\rho_A\otimes\rho_B+\frac{1}{2d_A}\sum_{\mu,\nu}
 C_{\mu\nu}\sigma_{A\mu}\otimes\sigma_{B\nu}\label{rab}\end{equation}
where $\rho_A=\frac{1}{d_A}(I_A+\bm{r}_A\cdot\bm{\sigma}_A)$,
$\rho_B=\frac{1}{2}(I_2+\bm{r}_B\cdot\bm{\sigma})$  are the reduced states of
$A$ and $B$, with $\bm{r}_A=\langle\bm{\sigma}_A\rangle$,
$\bm{r}_B=\langle\bm{\sigma}\rangle$,  and
\begin{equation}
C_{\mu\nu}=\langle \sigma_{A\mu}\otimes\sigma_{B\nu}\rangle-\langle\sigma_{A\mu}\rangle
\langle\sigma_{B\nu}\rangle\label{Ct}
\end{equation}
are  the elements of the {\it correlation tensor}, represented by the
$(d_A^2-1)\times 3$ matrix $C$.

We consider a local POVM measurement on the qubit $B$ based on rank one
operators $M_{\bm{k}B}= \sqrt{q_{\bm{k}}}P_{{\bm k}B}$, where
$P_{\bm{k}B}=\frac{1}{2}(I_2+\bm{k}\cdot\bm{\sigma})$, with $\bm{k}$ a unit
vector ($|\bm{k}|=1$), is the projector onto the pure qubit state with $\langle
\bm{\sigma}\rangle=\bm{k}$, and $\sum_{\bm{k}} q_{\bm{k}} P_{\bm{k}B}=I_2$. We
may then express the ensuing conditional entropy (\ref{Sfc}) as
\begin{equation}
S_f(A|B_{\bm{k}})=\sum_{\bm{k}}p_{\bm{k}}S_f(\rho_{A/\bm{k}})\label{Sfk}
\end{equation}
where
\begin{eqnarray}
\rho_{A/\bm{k}}&=&\rho_A+\frac{1}{d_A}(\frac{C\bm{k}}{1+\bm{r}_B\bm{k}})
\cdot\bm{\sigma}_A\\
p_{\bm{k}}&=&\frac{1}{2}q_{\bm{k}}(1+\bm{r}_B\cdot\bm{k})
\end{eqnarray}
are, respectively, the conditional post measurement state of $A$ after result
$\bm{k}$ and the probability of obtaining this result. The vector
$\bm{r}_{A/\bm{k}}$ characterizing the post-measurement state of $A$ is then
\begin{equation}\bm{r}_{A/\bm{k}}=\bm{r}_A+\frac{C\bm{k}}{1+\bm{r}_B\cdot\bm{k}}\,.
\label{rak}\end{equation}
For a standard projective spin measurement along direction $\bm{k}$
just vectors $\pm\bm{k}$ are to be considered in the previous sums,
with $q_{\pm\bm{k}}=1$.

While in the general case the eigenvalues of $\rho_{A/\bm{k}}$ are required for
the evaluation of (\ref{Sfk}), for the quadratic entropy (\ref{S2}) a closed
evaluation is directly feasible with Eq.\ (\ref{S2r}). For a standard
projective spin measurement along direction $\bm{k}$ we obtain \cite{GR.14}
\begin{eqnarray}
S_2(A|B_{\bm{k}})&=&S_2(\rho_A)-\Delta S_2(A|B_{\bm k})\,,\label{S2k}\\
\Delta S_2(A|B_{\bm{k}})&=&\frac{2}{d_A}\frac{|{C}\bm{k}|^2}{1-(\bm{r}_B
\cdot\bm{k})^2}=\frac{2}{d_A}\frac{\bm{k}^TC^TC\bm{k}}
{\bm{k}^T N_B\bm{k}}\,,\label{dS2k}
\end{eqnarray}
where $C^TC$ and $N_B=I-\bm{r}_B\bm{r}_B^T\,,$  are $3\times 3$ positive
semi-definite matrices. Eq. (\ref{dS2k}) is  non-negative and independent of
$\bm{r}_A$, and represents the average conditional purity gain due to the
measurement on $B$. Since Eq. (\ref{dS2k}) is a ratio of quadratic forms, the
direction $\bm{k}$ which leads to the {\it maximum} entropy decrease, i.e. to
the {\it minimum} conditional entropy,  can be obtained by solving the
generalized eigenvalue equation \cite{GR.14}
\begin{equation}
C^TC\bm{k}=\lambda N_B\bm{k}\,,\label{eig2}
\end{equation}
which implies ${\rm Det}[C^TC-\lambda N_B]=0$, and selecting the eigenvector
$\bm{k}$ associated with the largest eigenvalue $\lambda_{\rm max}$. This leads
to $\Delta S_2(A|B_{\bm k})\leq 2\lambda_{\rm max}/d_A$ $\forall$ $\bm{k}$,
i.e.,
\begin{equation}
\mathop{\rm Min}_{\bm{k}}S_2(A|B_{\bm{k}})=S_2(\rho_A)-\frac{2}{d_A}\lambda_{\rm max}\,.
\label{S2min}
\end{equation}
An important remark is that  generalized POVM
measurements on qubit $B$ {\it cannot decrease the projective
minimum (\ref{S2min})}   for this entropy \cite{GR.14}.

\begin{figure}[t]
 \includegraphics*[scale=.5]{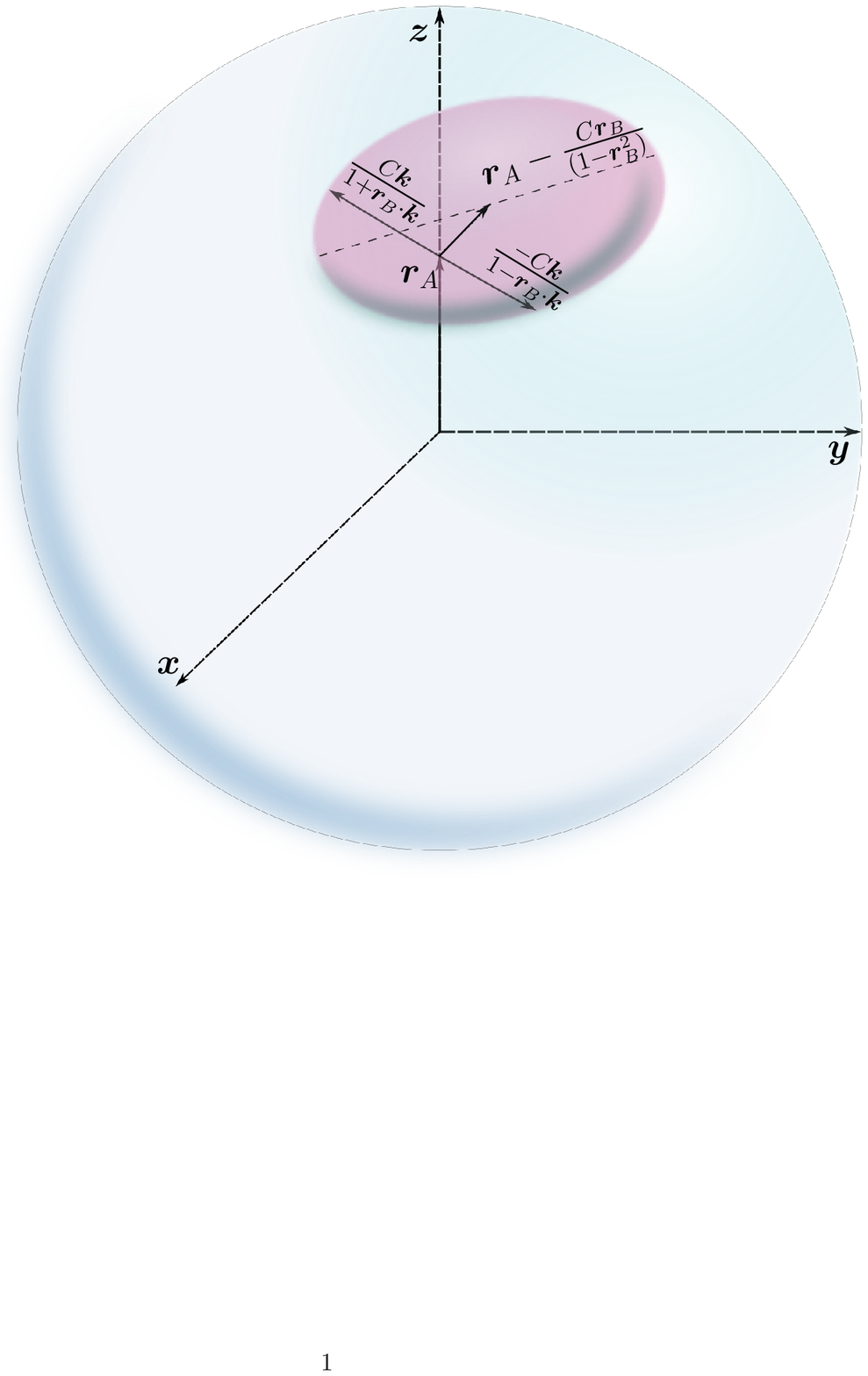}
\caption{The set of possible Bloch vectors $\bm{r}_{A/\bm{k}}$ of the
post-measurement state of qubit $A$ after a measurement on the qubit $B$ form
the correlation ellipsoid (from \cite{GR.14}). For a spin measurement along
direction $\bm{k}$ at qubit $B$, the vectors $\bm{r}_{A/\pm\bm{k}}$ in $A$ are
the endpoints of a chord running through $\bm{r}_A$. The optimizing  measurement 
determined by Eq.\ (\ref{eig2}) leads to $\delta\bm{r}_A$ parallel to the major axis of this 
ellipsoid.}   \label{f1}
\end{figure}

It is then  seen that the minimizing measurement is essentially determined by
the correlation tensor (\ref{Ct}), i.e., it is essentially a spin measurement
along the  direction of {\it maximum correlation}. We may also express
(\ref{dS2k}) as the  quadratic form $\Delta
S_2(A|B_{\bm{k}})=\frac{2}{d_A}\bm{k}_N^TC_N^TC_N\bm{k}_N$, where
$C_N=CN_B^{-1/2}$ and $\bm{k}_N=N_B^{1/2}\bm{k}/|N_B^{1/2}\bm{k}|$, and write
(\ref{eig2}) as $C_N^TC_N\bm{k}_N=\lambda\bm{k}_N$, which shows that
$\sqrt{\lambda_{\rm max}}$ is the {\it maximum singular value} of $C_N$. The
counterpart at $A$ of this equation  is $C_NC_N^T\bm{k}_A=\lambda\bm{k}_A$,
which has the same non-zero eigenvalues and provides a clear {\it geometric
picture:} As $\bm{k}$ is varied in the Bloch sphere of qubit $B$, the set of
post-measurement vectors (\ref{rak}) determining the post-measurement state of
$A$ form a three dimensional {\it correlation ellipsoid} on the $d_A^2-1$
dimensional space containing the vector $\bm{r}_{A/\bm{k}}$ \cite{GR.14} (see
Fig.\ \ref{f1} for the two-qubit case) whose  principal axes are precisely
determined as the eigenvectors $\bm{k}_A$ of the previous equation. Therefore,
the optimizing measurement of the quadratic entropy is  that leading to
$\delta\bm{r}_A=\bm{r}_A-\bm{r}_{A/\bm{k}}\propto C\bm{k}$ parallel to the major semi-axis of the
correlation ellipsoid (see \cite{GR.14} for more details).

While not strictly valid for other entropies, these results provide an
approximate  picture of the measurement minimizing the conditional entropy in
these systems, which will typically lie close to that minimizing the quadratic
entropy. In fact, all entropies $S_f(\rho)$ reduce essentially to the quadratic
entropy if $\rho$ is sufficiently close to maximum mixedness, as
$S_f(\frac{I}{d}+\delta\rho) \approx
S_f(\frac{I_d}{d})+\frac{1}{4}|f''(\frac{1}{d})|
 [(S_2(\frac{I}{d}+\delta\rho)-S_2(\frac{I_d}{d})]$
up to $O(\delta\rho^2)$. Moreover, for  a sufficiently small correlation
tensor, i.e. if $|\delta\bm{r}_A|=|\frac{C\bm{k}}{1\pm\bm{r}_B\bm{k}}|\ll 1$
$\forall$ $\bm{k}$, an expansion of the conditional entropy (\ref{Sfc})  up to
second order in $\delta\rho_A$ leads to \cite{GR.14}
\begin{equation}S_f(A|B_{\bm{k}})\approx S_f(\rho_A)-\frac{2}{d_A}
\frac{\bm{k}^TC^T\Lambda_f(\rho_A)C\bm{k}}{\bm{k}^T N_B\bm{k}}\end{equation}
where $\Lambda_f(\rho_A)$ is a scaled $(d_A^2-1)\times (d_A^2-1)$ Hessian matrix \cite{GR.14}, 
showing that in the present weakly correlated regime the effect 
of a general entropy is just to replace $C$ by the ``deformed'' correlation
tensor $C_f=\sqrt{\Lambda_f(\rho_A)}\,C$. Let us finally mention that 
the minimum generalized conditional entropy coincides with the associated generalized entanglement of formation between $A$ and a third system $C$  purifying the whole system 
\cite{KW.04,GR.14b}.

\subsection{Generalized information deficit}
\label{subsec2}
As mentioned in the introduction, several other measures of quantum correlations with  properties 
similar to those of the quantum discord have been considered. In particular, we have introduced in \cite{RCC.10,RCC.11} the {\it generalized information deficit} 
 \begin{equation}
 I_f^B(\rho_{AB})=\mathop{\rm Min}_{M_B}\, S_f(\rho'_{AB})-S_f(\rho_{AB})\,,
\label{If}
\end{equation}
where $\rho'_{AB}$ is the state  of the system after an unread local
measurement at $B$, Eq.\ (\ref{rhop}), and  the minimization is  over all
complete local projective measurements on B. Here $S_f(\rho)$ denotes a
generalized entropy. In the case of the von Neumann entropy $S(\rho)$, Eq.\
(\ref{If}) becomes the standard one-way information deficit
\cite{HHH.05,SKB.11,Mo.12}, which will be denoted as $I_1^B$. It can be
rewritten in terms of the relative entropy \cite{Ww.78,Ve.02}
$S(\rho||\rho')=-{\rm Tr}\,\rho (\log_2\rho'-\log_2\rho)$ as
\begin{equation} I^B_1(\rho_{AB})=\mathop{\rm Min}_{M_B} S(\rho'_{AB})-S(\rho_{AB})=
\mathop{\rm Min}_{M_B} S(\rho_{AB}||\rho'_{AB})\,. \label{IS}\end{equation}
Like the quantum discord, Eq.\ (\ref{If}) (and hence (\ref{IS})) is
non-negative if $S_f(\rho)$ is Schur-concave \cite{Bha.97}, due to the
majorization relation \cite{RCC.10,Ww.78} $\rho'_{AB}\prec\rho_{AB}$
satisfied by the post-measurement state (\ref{rhop}). Essentially, the
off-diagonal elements of $\rho_{AB}$ in the conditional product basis
$\{|i_{A}^j\rangle|j_B\rangle\}$ where $\rho'_{AB}$ is diagonal are lost in the
measurement, and Eq.\ (\ref{If}) is then a measure of the minimum information
loss under such measurement. It can also be considered as the minimum relative
entropy of coherence  \cite{BCP.14} in this type of basis. And it is a measure of the minimum entanglement between the measurement device and the system generated by a complete local measurement \cite{CCR.15}, with (\ref{IS}) representing the minimum distillable entanglement \cite{SKB.11,PG.11}. 

For strict concavity of $S_f$, Eq.\ (\ref{If}) vanishes only if $\rho_{AB}$ is
already of the semi-quantum post-measurement form (\ref{rhop}). And for pure
states  $\rho_{AB}=|\Psi_{AB}\rangle\langle\Psi_{AB}|$ it can be shown \cite{RCC.10} that it
reduces to the corresponding entanglement entropy:
\begin{equation}I_f^B(|\Psi_{AB}\rangle)=S_f(\rho_A)=S_f(\rho_B)\end{equation}
with (\ref{IS}) becoming the standard entanglement entropy like the quantum
discord. Nonetheless, unlike the latter (which in this case is minimized by a
complete measurement in any local basis) the  minimum of (\ref{If}) and
(\ref{IS}) for a pure state is always reached for a measurement in the basis of
$B$ which corresponds to the Schmidt decomposition of $|\Psi_{AB}\rangle$ (and
hence diagonalizes $\rho_B$) \cite{RCC.10}, 
already indicating a different behavior of the minimizing measurement.

As in the case of the conditional entropy, the use of generalized entropies
enables the possibility of using simple entropic forms like the quadratic
entropy (\ref{S2}) or the Tsallis entropies, in which case Eq.\ (\ref{If})
becomes $I_q^B(\rho_{AB})=\mathop{\rm Min}_{M_B}\frac{{\rm Tr}\,
(\rho^{q}_{AB}-{\rho'}_{AB}^{q})}{1-2^{1-q}}$. We may also consider the deficits
based on the Renyi entropies \cite{Ww.78} $S_{R_q}(\rho)=\frac{1}{1-q}\log_2 {\rm
Tr}\,\rho^q$, $q>0$ (just an increasing function of $S_q$),  which are given by  \cite{CCR.15}
\begin{equation}
I^{B}_{R_q}(\rho_{AB}) = \mathop{\rm Min}_{M_B} \frac{1}{1-q}
 \log_2 \frac{{\rm Tr}\,{\rho'}^{q}_{AB}}{{\rm Tr}\,\rho_{AB}^q}\,. \label{IR}
 \end{equation}
They approach the von Neumann information deficit (\ref{IS}) for
$q\rightarrow 1$ and likewise do not depend on the addition of an uncorrelated
ancilla to $A$ ($\rho_{AB}\rightarrow \rho_C\otimes \rho_{AB}$). Nonetheless, they
are just increasing functions of $I_q^B$ for fixed $\rho_{AB}$ and the
associated optimization problem is the same as that for $I_q^B$.

\subsection{Minimizing measurement and stationary conditions}
 \label{sec:3}
The determination of the minimizing measurement $M_B$ in (\ref{If}) is, like in
the case of the quantum discord, again a difficult problem in general. Complete
projective measurements at $B$ are determined by $d^2_B-d_B$ real parameters if
$B$ has  Hilbert space dimension $d_B$, growing then exponentially with the
number of components of $B$. Nevertheless, it can be shown  that the minimizing
measurement should fulfill the stationary condition \cite{RCC.11}
\begin{equation}
{\rm Tr}_{A}[f'(\rho'_{AB}),\rho_{AB}]=0 \,,
\label{stat}
\end{equation}
which leads to $d_B(d_B-1)$ real equations \cite{RCC.11,RMC.12}. In the quantum
discord (\ref{D1}), an additional term
$-[f'(\rho'_B),\rho_B]=[\log_2\rho'_B,\rho_B]$ is to be added in (\ref{stat})
for complete projective measurements \cite{RCC.11}.

Important differences between the measurements minimizing   $I_f^B(\rho_{AB})$
and $D(A|B)$ may arise, as previously mentioned for the case of pure states.
While for a general classically correlated state of the form (\ref{rhop}) the
minimum for {\it both} $D(A|B)$ and {\it all} $I_f^B(\rho_{AB})$ is attained
for a measurement in the local basis defined by the projectors $P_j^B$ (i.e.,
the pointer basis \cite{OZ.01,HV.01}), in the particular case of product states
$\rho_A\otimes \rho_B$, $D(A|B)$ ({\it but not $I_f^B(\rho_{AB})$}) becomes the
same for {\it any} $M_B$, as for such states $S(A|M_B)=S(A)$ $\forall$ $M_B$.
These differences will have important consequences in the results of the next
section, leading to a quite different response of the minimizing measurement to
the onset of quantum correlations. They reflect the fact that while in
$I_f^B(\rho_{AB})$ one is looking for the {\it least disturbing local
measurement}, such  that $\rho'_{AB}$ is as close as possible to $\rho_{AB}$,
in $D(A|B)$ the search is for the measurement in $B$ which makes the ensuing
conditional entropy smallest, i.e., by which one can learn the most about $A$,
which leads to those observables which are most correlated, as discussed
before.

These differences become apparent in the case of the quadratic entropy
(\ref{S2}), as an analytic evaluation of the associated deficit $I_2^B$ for
qudit-qubit systems becomes again feasible \cite{DVB.10,RCC.11}. In this case
Eq.\ (\ref{If}) becomes just a purity difference, $I_2^B(\rho_{AB})=
2\mathop{\rm Min}_{M_B}{\rm Tr}\,
(\rho^{2}_{AB}-{\rho'}_{AB}^{\,2})=2\mathop{\rm Min}_{\rho'_{AB}}||\rho_{AB}
-\rho'_{AB}||^2$, where $||O||^2={\rm Tr}\,O^\dagger O$ and the last
minimization can be extended to any state of the general form (\ref{rhop}).
Through the last expression it is seen that it is then proportional to the
geometric discord \cite{DVB.10,Mo.12}, defined as the closest squared
Hilbert-Schmidt distance between $\rho_{AB}$ and a state of the form
(\ref{rhop}).  For pure states $I_2^B$ becomes the squared concurrence
$C^2_{AB}$ \cite{HW.97}, which for such states is just the quadratic entropy of
any of the subsystems \cite{Ca.03}. While as a measure it does not comply, due
to the lack of additivity, with all the properties satisfied by the quantum
discord or the von Neumann based information deficit,  it has the advantage of
enabling a simple analytic evaluation in qudit-qubit systems and admitting
through its relation with the purity a more direct experimental access
\cite{F.02,F.02b,F.02c}. Moreover, the optimizing measurement will be the same measurement
as that minimizing the associated Renyi deficit $I^{B}_{R_2}(\rho_{AB})$.

Writing  again a general $\rho_{AB}$ of a qudit-qubit system in the form
  (\ref{rab}),
it can be shown that for a projective spin measurement at $B$ along
direction $\bm{k}$, the quadratic information loss becomes
\cite{DVB.10,RCC.11}
\begin{equation}
I^B_2(\bm{k})={\textstyle\frac{1}{d_A}}(||\bm{r}_B||^2+||J||^2-\bm{k}^T M_2 \bm{k})
\label{IB2xx}
\end{equation}
where $M_2$ is the positive semi-definite matrix
\begin{equation}M_2=\bm{r}_B\bm{r}_B^T+J^TJ\,,\label{M2}\end{equation}
with $J=C+\bm{r}_A\bm{r}_B^T$, i.e., $J_{\mu\nu}=\langle
\sigma_{A\mu}\otimes\sigma_{\nu}\rangle$. Minimization of $I^B_2(\bm{k})$ leads
then to the standard eigenvalue equation $M_2\bm{k}=\lambda \bm{k}$, implying
$I_2^B(\rho_{AB})={\textstyle\frac{1}{d_A}}({\rm tr}\,M_2-\lambda_{\rm max})$,
with $\lambda_{\rm max}$ the largest eigenvalue of $M_2$ and the minimizing
$\bm{k}$  the associated eigenvector. Such direction will not necessarily
coincide with that minimizing the quadratic conditional entropy, as the latter
is determined essentially by the correlation tensor $C$ while the present one
by the tensor $J$ and $\bm{r}_B$.  While coinciding in some regimes (they
become identical if $\bm{r}_B=\bm{0}$, i.e., $\rho_B$ maximally mixed, in which
case $J=C$), they can deviate considerably in others, as will be explicitly
shown in the next section. In fact, a transition in  the least disturbing
measurement direction $\bm{k}$ from  the main eigenvector of $J^T J$ to the
direction of $\bm{r}_B$ can be expected as $J$ decreases, which may not imply a
concomitant change in the main  eigenvector of (\ref{eig2}).  A closed expression
for the minimum of $I_3^B(\rho_{AB})$ can also be obtained \cite{RCC.11}. We
finally note that for a general qubit-qubit state and entropy $S_f$, the
stationary condition (\ref{stat}) becomes explicitly
\begin{equation}
 (\alpha_1\bm{r}_B+\alpha_2 J^T\bm{r}_A+\alpha_3J^TJ)\bm{k}=\lambda\bm{k}
\label{stat2}
\end{equation}
which represents a non-linear eigenvalue equation since the coefficients
$\alpha_i$ depend on $f'(\rho'_{AB})$ and hence on $\bm{k}$ \cite{RCC.11}.
Again, the prominent role of $J^TJ$ is clearly evident.

\section{Results in spin chains\label{sec3}}

We now consider the correlations of spin pairs in the ground
state (GS) of finite spin $1/2$ arrays interacting through $XY$ type Heisenberg
couplings and immersed in a magnetic field $\bm{h}$. The Hamiltonian reads
\begin{equation}
H=-\sum_{i} \bm{h}\cdot\bm{S}_i - {\textstyle\frac{1}{2}} \sum_{i\neq j,\mu=x,y} J_\mu^{ij}
S^\mu_i S^\mu_j\,, \label{H}
\end{equation}
where $i,j$ label the sites in the array and $S_i^\mu$ the spin
components at site $i$.

In the transverse case $\bm{h}=(0,0,h_z)$, the Hamiltonian commutes with the
{\it $S_z$ spin parity} $P_z=e^{i\pi\sum_i (S_{i}^z+1/2)}=\prod_i (-2S_{i}^z)$,
implying  that the exact GS will have a definite parity if non-degenerate.  In
particular, in finite chains of $N$ spins with first neighbor couplings and
anisotropy $\chi=J_y/J_x\in (0,1]$, the exact GS, which can be analytically obtained 
through the Jordan-Wigner fermionization \cite{LSM.61,RCM.08}, will exhibit $N/2$ parity
transitions as the field $h_z$ increases from $0$, where the lowest levels of
each parity cross, the last one at the {\it transverse factorizing field}
\cite{RCM.08} $h_{zs}=\sqrt{J_yJ_x}=J_x\sqrt{\chi}$. These transitions are
reminiscent of the $N/2$ magnetization transitions of the $XX$ case $\chi=1$
\cite{CCR.13,CR.07}, where $H$ commutes with the $z$ component of the total
spin $S^z=\sum_i S^z_i$.

\begin{figure}[t]
\includegraphics*[scale=.5]{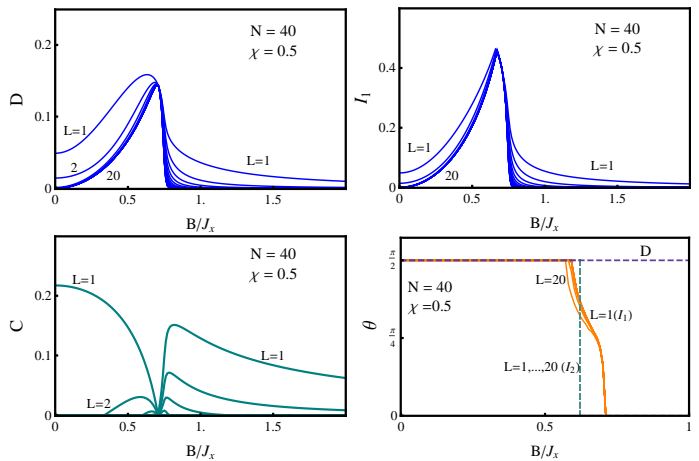}
\caption{The quantum discord (top left), the information deficit (top right),
the concurrence (bottom left) and the angle $\theta$ determining the minimizing
spin  measurement of the first two quantities (bottom right) for reduced states of spin
pairs in the exact ground state of a spin $1/2$ chain with first neighbor
anisotropic $XY$ coupling as a function of the transverse field ($B=h_z$). $L$ indicates
the separation between the spins of the pair  ($L=1$ denotes first neighbors)
while $N$ is the number of spins and $\chi=J_y/J_x$ the anisotropy. The angle
$\theta$ is that formed between the measurement direction and the $z$ axis in
the $xz$ plane, which is constant for $D$ ($\theta=\pi/2$) but experiences an 
$x\rightarrow z$ transition in the information deficits $I_1$ (solid line) and
$I_2$ (dashed line), which is sharp in the latter.  All quantities reach full
range at the factorizing field $B=J_x\sqrt{\chi}$, with common $L$-independent
limits, which are negligible in the case of the  concurrence but finite for the
discord and information deficit.} \label{f2}
\end{figure}

At the factorizing field, the two crossing states generate a two dimensional GS
subspace which is spanned, remarkably, by {\it completely separable} ground
states $|\Theta\rangle=|\theta,\ldots,\theta\rangle$ and
$|-\Theta\rangle=P_z|\Theta\rangle=|-\theta,\ldots,-\theta\rangle$ in the
ferromagnetic case $J_x>0$, where $|\theta\rangle=e^{-\imath\theta
S_y}|\downarrow\rangle$ is the  single spin state forming an angle $\theta$
with the $-z$ direction and $\cos\theta=h_{sz}/J_x=\sqrt{\chi}$.  Hence, at
this point the system possesses  two completely separable parity breaking
degenerate ground states. Yet, the exact GS side-limits at this point are
provided by the definite parity combinations \cite{RCM.08}
\begin{equation}|\Theta_{\pm}\rangle=\frac{|\Theta\rangle\pm|-\Theta\rangle}
{\sqrt{2(1\pm\langle -\Theta|\Theta\rangle)}}\,\label{stt}\end{equation}
approached for $h_z\rightarrow h_{zs}^{\pm}$, which are {\it entangled} states.
They  lead to  {\it common reduced states $\rho_{\theta\pm}$ for  any  spin
pair $i\neq j$} \cite{RCM.08,CRC.10},  becoming both identical with
$\rho_\theta=(|\theta\theta\rangle\langle\theta\theta|
+|-\theta-\theta\rangle\langle-\theta-\theta|)/2$ if the overlap $\langle
-\Theta|\Theta\rangle=\cos^N\theta$ is neglected (it is negligible if $N$ and
$\theta$ are not too small). This is a separable mixed state, therefore leading
to a zero concurrence (and hence zero entanglement of formation \cite{HW.97})
for any pair, as seen in the bottom left panel of Fig.\ \ref{f2}. The
concurrence actually approaches small common side limits
$C_{\pm}=\frac{\chi^{n/2-1}(1-\chi)}{1\pm\chi^{n/2}}$ if the overlap is
preserved \cite{RCM.08, RCM.09}, not appreciable in the scale of fig.\
\ref{f2}.

However, $\rho_\theta$ is a discordant state for $\theta\in(0,\pi/2)$, leading  to appreciable {\it
finite limits} of the quantum discord $D$ and the information deficit
$I_1$ at the factorizing field, as seen in the top panels of Fig.\ \ref{f2}.
These limits can be analytically determined from the previous expression for
$\rho_\theta$  \cite{CRC.10,CCR.15} and are independent of the separation $L$.
Moreover, these quantities actually attain their maximum values in the vicinity
of this point, remaining appreciable for all $h_z<h_{zs}$, since in this sector
the reduced state of any pair in the exact  GS will be essentially
$\rho_{\theta}$ (with a field-dependent $\theta$) plus smaller  corrections.
Let us note that in the cyclic chain considered, the reduced pair states
$\rho_{ij}$ depend just on the separation $L=|i-j|$, implying $D(i|j)=D(j|i)=D$
and $I^i_f(\rho_{ij})=I^j_f(\rho_{ij})=I_f$ $\forall$ $i\neq j$.

These results remain strictly valid for arbitrary range couplings with a common
anisotropy $\chi=J^{ij}_y/J^{ij}_x$  in the ferromagnetic case $J^{ij}_x>0$
\cite{RCM.08}, including dimer-type chains \cite{CRM.10},  since they also
exhibit a factorizing field with the same factorized states. They hold as well
in the antiferromagnetic case $J_x<0$ for first neighbor couplings in a spin
chain, since for a transverse field it can be mapped to the ferromagnetic case
by a local rotation at even sites, which leads to
$|\theta\theta\ldots\rangle\rightarrow|\theta,-\theta,\ldots\rangle$ in the
factorized states and in $\rho_{\theta}$. The same reduced state $\rho_\theta$
also follows from the mixture $\frac{1}{2}(|\Theta_+\rangle\langle
\Theta_+|+|\Theta_-\rangle\langle\Theta_-|)$ if the overlap is neglected, which
is  the exact $T\rightarrow 0^+$ limit at $h_{zs}$ of the thermal state
$\propto\exp[-H/kT]$. We remark finally that in the thermodynamic limit
$N\rightarrow\infty$, the lowest states for each parity become degenerate for
$h_{z}<h_{c}=\frac{J_x+J_y}{2}$, so GS correlations actually depend on the
choice of GS, the present results applying for the definite parity choice.

As seen in Fig.\ \ref{f2}, although the quantum discord $D$ and the information
deficit $I_1$ exhibit a similar qualitative behavior,  $I_1$ shows a more
pronounced maximum in comparison with $D$. This feature reflects the transition
in the orientation $\bm{k}$ of the local  spin measurement minimizing  $I_1$ as
the field increases,  which is absent in the quantum discord.  This effect can
be understood from the expressions (\ref{IB2xx})--(\ref{M2}) for the quadratic
deficit $I_2$, which lead to a sharp $x\rightarrow z$ transition in the
optimizing $\bm{k}$ for all separations as the maximum eigenvalue of $M_2$
shifts from  that associated with $\bm{k}=\bm{e}_x$ to that for
$\bm{k}=\bm{e}_z$ as the transverse  field increases \cite{RCC.11,CCR.15}. In
the case of $I_1$ such sharp transition is smoothed, as seen in the bottom
right panel of Fig.\ \ref{f2}, with $\theta$ covering all  intermediate values
in a narrow field interval centered at the $I_2$ measurement transition. In
contrast, the quantum discord prefers a spin measurement (we consider here
projective spin measurements) along the $x$ axis for {\it all} transverse
fields,   for any separation $L$, following the strongest correlation
\cite{GR.14,GR.14b}, which is along $x$ for $|J_x|>|J_y|$. This is precisely
the same measurement minimizing the quadratic conditional  entropy, determined
by Eqs.\ (\ref{dS2k})--(\ref{S2min}), since the largest eigenvalue in (\ref{eig2}) of the
contracted correlation matrix $C^TC$ corresponds to $\bm{k}$ along the $x$ axis
for the present anisotropic $XY$ coupling $\forall$ $h_z$ \cite{GR.14,GR.14b}.

The  measurement transitions of the information deficit reflect, on the other hand, the qualitative
change undergone by the reduced state of the pair (essentially by its dominant
eigenstate) as the field increases \cite{CCR.15}. The same $x\rightarrow z$
transition in $I_1$ is found in the $XX$ case, where it reflects  the
transition  in the dominant eigenstate of the reduced state of the pair from a
Bell state
$\frac{|\uparrow\downarrow\rangle+|\downarrow\uparrow\rangle}{\sqrt{2}}$ to the
aligned state $|\downarrow\downarrow\rangle$ as the transverse field increases
\cite{CCR.13}. The main correlation in $C^TC$ stays, however, along the $x$
axis. And in spin $1$ systems, while the local optimizing measurements become
more complex (they are not standard spin measurements), a similar transition
pattern is observed in the measurement minimizing the information deficit
\cite{RMC.12}.

\begin{figure}[t]
\includegraphics[scale=.575]{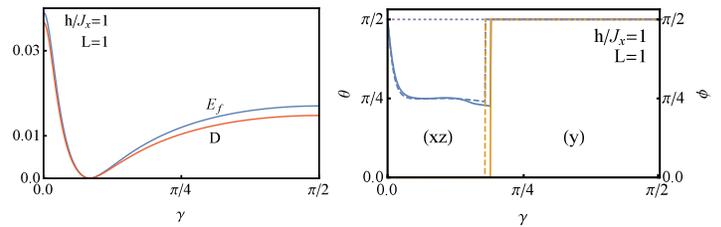}
\caption{The quantum discord $D$ and the entanglement of formation $E_f$ (left
panel), and the angles $\theta$, $\phi$ determining the minimizing spin
measurement direction
$\bm{k}=(\sin\theta\cos\phi,\sin\theta\sin\phi,\cos\theta)$ for $D$ (right
panel, solid lines), for a first neighbor pair in a $XY$ spin chain with
$\chi=0.5$, and a non-transverse field $\bm{h}$ in the $xz$ plane, as a function
of the angle $\gamma$ it forms with the $z$ axis, for $|\bm{h}|=J_x$. Here both
$D$ and $E_f$ vanish at the factorizing field due to the non-degeneracy of the
factorized ground state. A transition from the $xz$ plane to the $y$ axis takes
place in the minimizing projective measurement as $\gamma$ increases,  which
can be predicted through the measurement optimizing the quadratic conditional
entropy (dashed lines in right panel).} \label{f3}
\end{figure}

In Fig.\ \ref{f3} we depict illustrative results for a non-transverse field
$\bm{h}=(h_x,0,h_z)$ in the $xz$ plane, for an $XY$ chain with coupling
anisotropy $\chi=0.5$ and small spin number $N=8$. As recently shown \cite{CRC.15}, such
chains also exhibit a non-transverse GS factorizing field in the $xz$ plane,
whose magnitude is given by
\begin{equation}|\bm{h}_{s}|=\frac{h_{zs}\sin\theta}{\sin(\theta-\gamma)}
\end{equation}
where $h_{zs}=J_x\cos\theta$ is the transverse factorizing field, with
$\cos\theta=\sqrt{\chi}$ and $\gamma<\theta$  the angle formed by the field
with the $z$ axis. In contrast with the transverse case, such field is now
associated with a {\it non-degenerate} separable GS
$|\Theta\rangle=|\theta\theta\ldots\rangle$, as parity symmetry no longer
holds. It is then seen that both $D$ and the entanglement of formation $E_f$
exactly vanish at $\bm{h}_s$, being now smaller than in the previous case since
the GS  no longer has parity symmetry. Moreover, for first neighbors the
reduced pair state  is much less mixed than before,  and hence $E_f$ and $D$
have similar values, with  $E_f$  slightly larger than $D$, as also occurs
for strong transverse fields \cite{CRC.10}. For second and more distant
neighbors, the behavior of $D$ is qualitatively similar but  becomes smaller (and 
larger than $E_f$). It should be remarked that $E_f$ (and also $D$, $I_f$)
continues to exhibit {\it long range} in the vicinity of the non-transverse factorizing field
\cite{CRC.15}.

In addition, the quantum discord now also exhibits a measurement transition if 
the field is not too small, from the $xz$ plane to the $y$ axis ($\theta=\phi=\pi/2$) as the field rotates
in the $xz$ plane from the $z$ axis to the $x$ axis. This transition can be
understood through the quadratic conditional entropy, as the maximum eigenvalue
of the contracted correlation tensor $C^TC$ in (\ref{eig2}) jumps from  the
$xz$ block to the $y$ block as the field is rotated, following the main
correlation.  As verified in the right panel, the measurement minimizing the
quadratic conditional entropy lies very close to that minimizing the von
Neumann based quantum discord. In contrast, even though the information deficit
(not shown) still exhibits a behavior similar to that of $D$, the associated
minimizing measurement tends to align with the field for strong $|\bm{h}|$,  
deviating again considerably from that minimizing the quantum discord.

\section{Conclusions \label{sec4}}
We have first described a consistent extension to general concave entropic forms of the measurement dependent von Neumann conditional entropy for bipartite quantum systems. This extension, while providing a general characterization of the average information gain after such measurement, enables the use of simple entropic forms like the quadratic entropy, for which a closed evaluation of the minimizing measurement (leading to maximum purity gain) in terms of the correlation tensor becomes feasible for  general states of qudit-qubit systems. Such solution  admits a simple geometrical picture and allows to capture the main features of the projective measurement minimizing the quantum discord, which is then seen to follow essentially the direction of maximum correlation. In contrast, 
that minimizing the information deficit is essentially a least disturbing local measurement,
and can then exhibit significant differences with the latter. The entropic generalization of the one way information deficit was also described, and for the quadratic entropy a 
closed evaluation for qudit-qubit states becomes again feasible, which allows to identify the
previous  differences.

When considered in spin pairs immersed in finite $XY$ spin chains, both quantities, discord and information deficit, exhibit similar trends although with significant differences in the behavior 
of their optimizing measurements, which can be understood and predicted with the closed evaluations for the quadratic case. For transverse fields, these quantities exhibit appreciable  values and long range for fields $h<h_c$ in the exact definite parity ground state, reaching full range and becoming independent of the pair separation in the vicinity of the factorizing field. A measurement transition takes place in the information deficit, which is absent in the quantum discord. 
In contrast, for non-transverse factorizing fields parity symmetry is broken and 
these quantities become smaller, strictly vanishing at factorization.  Measurement transitions can occur in both quantities. 

\section*{acknowledgement}
The authors acknowledge support from CONICET (NG,NC,MC) and CIC (RR) of Argentina.

\end{document}